\documentclass[a4paper,fleqn]{cas-dc}

\usepackage[numbers]{natbib}
\usepackage{stfloats} % en el preámbulo
\usepackage[ruled,vlined,linesnumbered]{algorithm2e}
\usepackage{booktabs}
\usepackage{multirow}
\usepackage{tabularx}
\usepackage[table]{xcolor}
\usepackage{makecell} % para saltos de línea en encabezados
\usepackage{amsthm}

\def\tsc#1{\csdef{#1}{\textsc{\lowercase{#1}}\xspace}}
\tsc{WGM}
\tsc{QE}
\tsc{EP}
\tsc{PMS}
\tsc{BEC}
\tsc{DE}
\newtheorem*{rmk}{Remark}
\begin{document}
\let\WriteBookmarks\relax
\def\floatpagepagefraction{1}
\def\textpagefraction{.001}

\shorttitle{Forecasting Seasonal Peaks of Pediatric Respiratory Infections Using an Alert-Based Model.}

\shortauthors{Henríquez et~al.}

\title [mode = title]{Forecasting Seasonal Peaks of Pediatric Respiratory Infections Using an Alert-Based Model Combining SIR Dynamics and Historical Trends in Santiago, Chile}                      
%\tnotemark[1,2]

%\tnotetext[1]{This document is the results of the research project funded by the National Science Foundation.}

%\tnotetext[2]{The second title footnote which is a longer text matter to fill through the whole text width and overflow into another line in the footnotes area of the first page.}

\author[1,2]{Gloria Henríquez}%[type=editor,
                        %auid=000,bioid=1,
                        %prefix=Sir,
                        %role=Researcher,
                        %orcid=0000-0001-7511-2910]

\cormark[1]
\ead{ghenriquez@dim.uchile.cl}


%\ead{cvr_1@tug.org.in}

%\ead[url]{www.cvr.cc, cvr@sayahna.org}

%\credit{Conceptualization of this study, Methodology, Software}

\affiliation[1]{organization={Center for Mathematical Modeling, Universidad de Chile and CNRS (IRL 2807)},
    %addressline={Beauchef 851, North Building, 7th Floor}, 
    city={Santiago},
    % citysep={}, % Uncomment if no comma needed between city and postcode
    %postcode={8370456}, 
    country={Chile}}

\affiliation[2]{organization={Institute of Biomedical Sciences, Universidad de Chile},
    %addressline={Beauchef 851, North Building, 7th Floor}, 
    city={Santiago},
    % citysep={}, % Uncomment if no comma needed between city and postcode
    %postcode={8370456}, 
    country={Chile}}

\author[1]{Jhoan Báez}%[style=chinese]

\author[1]{Víctor Riquelme}%[%
   %role=Co-ordinator,
   %suffix=Jr,
%\ead{cvr3@sayahna.org}
%\ead[URL]{www.sayahna.org}

%\credit{Data curation, Writing - Original draft preparation}

%\affiliation[2]{organization={Sayahna Foundation},
    %city={Jagathy},
    % citysep={}, % Uncomment if no comma needed between city and postcode
    %postcode={695014}, 
    %state={Trivandrum},
    %country={India}}


%\cormark[2]
%\fnmark[1,3]
%\ead{rishi@stmdocs.in}
%\ead[URL]{www.stmdocs.in}
%\author[?]{Samuel Marin}
%\affiliation[3]{organization={STM Document Engineering Pvt Ltd.},
    %addressline={Mepukada}, 
    %city={Malayinkil},
    % citysep={}, % Uncomment if no comma needed between city and postcode
    %postcode={695571}, 
    %state={Trivandrum},
    %country={India}}

\author[3]{Pedro Gajardo}
\affiliation[3]{organization={Departamento de Matemática, Universidad Técnica Federico Santa María},
%Hospital Dr. Luis Calvo Mackenna},
    %addressline={Av. Antonio Varas 360}, 
    city={Valparaíso},
    % citysep={}, % Uncomment if no comma needed between city and postcode
    %postcode={7500539}, 
    country={Chile}}

\author[4]{Michel Royer}
\affiliation[4]{organization={Dr. Luis Calvo Mackenna Hospital},
    %addressline={Dr. Carlos Lorca Tobar 999}, 
    city={Santiago},
    % citysep={}, % Uncomment if no comma needed between city and postcode
    %postcode={8380456}, 
    country={Chile}}

\author[1,5]{Héctor Ramírez}
\affiliation[5]{organization={Department of Mathematical Engineering, Universidad de Chile},
    %addressline={Beauchef 850}, 
    city={Santiago},
    % citysep={}, % Uncomment if no comma needed between city and postcode
    %postcode={8370456}, 
    country={Chile}}
\cortext[cor1]{Corresponding author at: University of Chile,
Beauchef 851, North Building, office 709, Santiago, Chile. }

%\cortext[cor2]{Principal corresponding author}

%\fntext[fn1]{This is the first author footnote. but is common to third author as well.}
%\fntext[fn2]{Another author footnote, this is a very long footnote and it should be a really long footnote. But this footnote is not yet  sufficiently long enough to make two lines of footnote text.}

%\nonumnote{This note has no numbers. In this work we demonstrate $a_b$ the formation Y\_1 of a new type of polariton on the interface  between a cuprous oxide slab and a polystyrene micro-sphere placed  on the slab. }

\begin{abstract}
Acute respiratory infections (ARI) are a major cause of pediatric hospitalization in Chile, producing marked winter increases in demand that challenge hospital planning. This study presents an alert-based forecasting model to predict the timing and magnitude of ARI hospitalization peaks in Santiago. The approach integrates a seasonal SIR model with a historical mobile predictor, activated by a derivative-based alert system that detects early epidemic growth. Daily hospitalization data from DEIS were smoothed using a 15-day moving average and Savitzky–Golay filtering, and parameters were estimated using a penalized loss function to reduce sensitivity to noise. Retrospective evaluation and real-world implementation in major Santiago pediatric hospitals during 2023 and 2024 show that peak date can be anticipated about one month before the event and predicted with high accuracy two weeks in advance. Peak magnitude becomes informative roughly ten days before the peak and stabilizes one week prior. The model provides a practical and interpretable tool for hospital preparedness.
\end{abstract}

% \includegraphics{figs/grabs.pdf}

%\item Research highlights item 1
%\item Research highlights item 2
%\item Research highlights item 3

% Keywords

\maketitle
%------------------------
\section{Introduction}
%------------------------

Acute respiratory infections (ARI) are among the leading causes of pediatric morbidity and hospitalization in Chile, exhibiting strong seasonality with recurrent winter epidemics \cite{achebak2023, suryadevara2021}. In the Metropolitan Region of Santiago, epidemic activity typically rises between May and September, driven by the seasonal circulation of respiratory viruses, environmental factors, and exposure to pollutants \cite{bulla1978}. These annual peaks exert severe pressure on the healthcare system: pediatric hospitals face bed shortages, emergency departments exceed their capacity, and routine care becomes disrupted during high-demand periods \cite{goic2021}. The burden is particularly substantial among infants, preschool-aged children, and school-aged populations. Respiratory syncytial virus (RSV) remains the principal etiological agent, followed by parainfluenza and influenza viruses, and, since 2020, SARS-CoV-2 has contributed additional complexity to epidemic dynamics \cite{cuadrado}.

Despite the clear seasonality of ARI, the {\bf timing and magnitude} of epidemic peaks vary substantially from year to year due to population immunity, viral co-circulation, climatic variability, and changes in mobility and contact patterns. This interannual unpredictability presents a major challenge for operational planning. In Chile, preparedness measures—including the nationwide “Winter Campaign” \cite{informeinvierno}—are traditionally guided by historical trends, routine surveillance, and expert judgment. However, this qualitative approach can result in premature or delayed allocation of critical resources such as beds, clinical staff, and equipment. For hospital administrators, the difference of a few weeks in peak timing can significantly affect patient outcomes and system efficiency.

Motivated by these type of operational needs, forecasting models have become an essential tool in public health decision-making. Internationally, several modeling approaches have been explored, including statistical time-series models \cite{almeida2022, becerra2020}, mechanistic epidemiological models \cite{ardabili2020, cadoni2020, fome2023, hota2021, piazzola2021, reis2016, weber2001}, and machine learning approaches \cite{albrecht2024, arias2023, casalegno2023, khatri2017early, lu2021, peng2020}. More recently, hybrid and ensemble methodologies \cite{reis2019} have gained traction, as they balance the interpretability of mechanistic modeling with the adaptability of data-driven techniques. However, most existing approaches focus on short-term prediction horizons, rely on retrospective data, or are not designed to integrate directly into day-to-day operational workflows within healthcare systems.

In this study, we propose an {\bf alert-based predictive framework aimed at forecasting, with adequate lead time, the timing and magnitude of peaks in pediatric acute respiratory infection (ARI) hospitalizations in Santiago, Chile}. The framework is developed with a strong focus on operational usability, supporting hospital managers in decision-making throughout the ascending phase of the epidemic.
The methodology yields an alert system that identifies key turning points in the epidemic curve—namely, the onset of growth, the point of maximum acceleration, and the inflection point—enabling early prediction and dynamic updating as new data arrive.
%A central aspect of our methodology is an alert system that detects key turning points in the epidemic curve—namely the onset of growth, the moment of maximum acceleration, and the inflection point—allowing predictions to begin early and evolve dynamically as new data arrive.

The proposed model follows a dual-prediction strategy:

\begin{itemize}
    \item the peak date prediction is required well in advance (ideally more than one month before the peak), as it informs staffing, bed allocation, and inter-hospital coordination;
    \item the peak magnitude prediction becomes reliable later in the season (typically one to two weeks before the peak), when the epidemic trajectory is more clearly defined.
\end{itemize}

To meet these two operational horizons, the framework integrates a seasonal SIR mechanistic model with a mobile historical prediction, combining them through a dynamically adjusted weighting function. 
This hybrid combination allows early forecasts to benefit from historical seasonal patterns, while progressively shifting weight towards the mechanistic model as the epidemic curve approaches its peak. 
%
%This ensemble structure allows early forecasts to benefit from historical seasonal patterns, while progressively shifting weight toward the mechanistic model as the epidemic curve approaches its peak. 
The result is a predictive system that is both interpretable and adaptive, capable of producing consistent daily updates with meaningful uncertainty ranges.

Daily hospitalization data were obtained from the official health records maintained by the {\em Department of Health Statistics and Information (DEIS)} \cite{DEIS}, which provides nationwide information on emergency care events. For this study, we primarily focused on the three major pediatric referral hospitals in the Santiago Metropolitan Region, which exclusively provide care to children and concentrate the highest pediatric respiratory burden: {\em Dr.~Luis Calvo Mackenna Children’s Hospital} (HLCM), {\em Dr.~Exequiel González Cortés Hospital} (HEGC), and {\em Dr.~Roberto del Río Children’s Clinical Hospital} (HRDR).

\begin{figure}
	\centering
		\includegraphics[scale=.49]{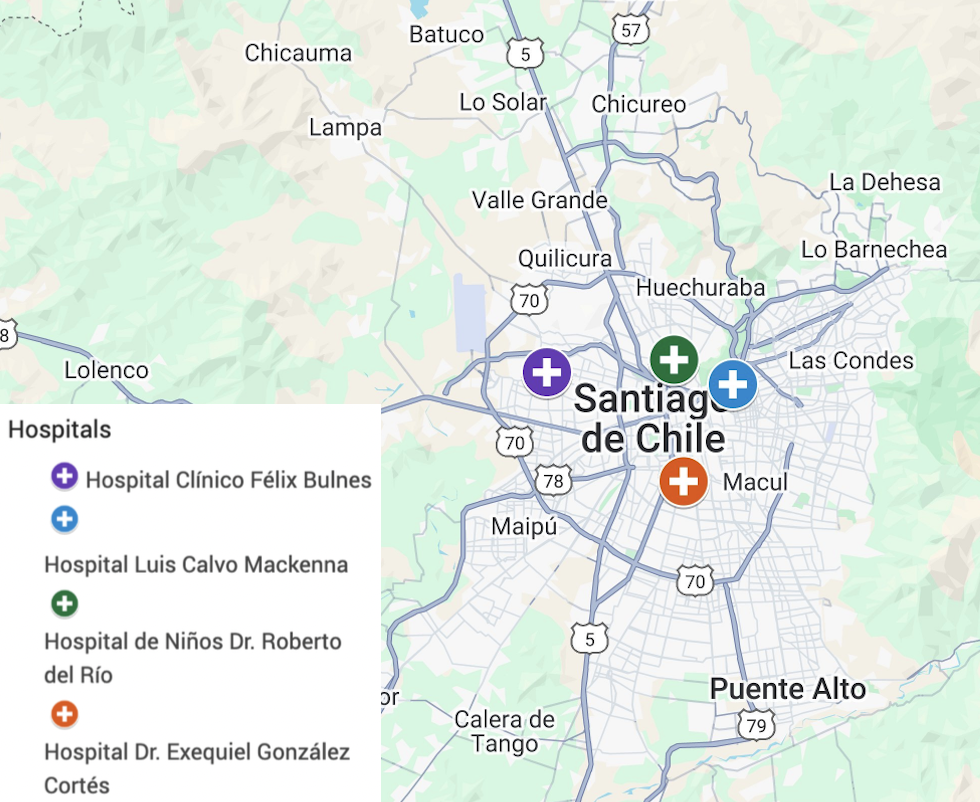}
	\caption{Geographic location of the four study hospitals in Santiago, Chile. The map highlights HFB (purple), HLCM (blue), HEGC (orange), and HDRD (green). Map created with Google My Maps.}
	\label{FIG:mapa}
\end{figure}

In addition, {\em Dr.~Félix Bulnes Cerda Hospital} (HFB) was included as a fourth study site due to its substantial pediatric hospitalization load, allowing for broader comparative analysis across different institutional profiles. The geographic locations of these institutions are shown in Figure~\ref{FIG:mapa}.

% aqui se comenta lo de fijar el hospital y de la omision de 2020 y 2021
All predictions were generated at the hospital level, as the proposed model requires hospital-specific calibration to account for differences in patient populations, historical demand patterns, and local epidemiological dynamics (see Appendix~\ref{appendix} for details). After filtering to the pediatric age range (patients younger than 15 years), the resulting dataset captures the temporal behavior of ARI-related hospitalizations across a representative segment of Chile’s pediatric healthcare system. Data from the years 2020 and 2021 were excluded from the historical analysis due to the disruptions introduced by the COVID-19 pandemic, which substantially altered hospitalization patterns and seasonal dynamics.

The main contributions of this work are the following:

\begin{itemize}
\item \textbf{Operational forecasting:} We develop a locally adapted model capable of predicting, in real time, both the peak date and peak magnitude of pediatric ARI hospitalizations.
\item \textbf{Alert-based methodological integration:} We propose a novel alert system that activates the forecasting pipeline and mediates the combination of epidemiological and historical components, providing early and robust predictions.
\item \textbf{Practical validation:} We evaluate the model across multiple historical years, demonstrating its applicability and usefulness for healthcare planning and resource management in Chile.
\end{itemize}

It is worth noting that, while this study focuses on the Chilean case, the proposed methodology is designed to be applicable to other contexts with similar epidemiological dynamics and data availability.

The remainder of this paper is organized as follows.
Section~\ref{model} describes the alert system and the predictive model.
Section~\ref{evaluation} presents an evaluation of forecasting performance across several epidemic seasons.
Section~\ref{discusion} discusses implications for health system operations and limitations of the current approach.
Section~\ref{conclusion} summarizes the main findings and outlines directions for future work.
Details of algorithmic implementation and parameter tuning are provided in Appendix~\ref{appendix}.

%%%%%%%%%%%%%%%%%%%%%%%%%%%%%%%%%%%%%%%%%%%%%%%%%%%%%%%%%%%%%%%%%%%%%%%%%%%%%%%%%%%%%%%%%%%%%%%%%%%%%%%%%

\section{Alert-Based Predictive Model}\label{model}

The proposed predictive framework is designed to forecast two key characteristics of the annual winter peak of pediatric acute respiratory infections (ARIs) in Santiago, Chile:

\begin{itemize}
\item the peak date of daily hospital admissions, denoted by $\hat{\mathbf{t}}$;
\item the peak magnitude, or expected maximum daily hospitalization demand, denoted by $\hat{\mathbf{h}}$;
\end{itemize}

Operationally, Chilean pediatric hospitals require early identification of peak timing—ideally more than one month in advance—to allocate staff, schedule beds, and anticipate surges in clinical demand. Conversely, the peak magnitude prediction is operationally relevant with a shorter lead time (one to two weeks), as resource planning can leverage historical patterns and the evolving slope of the epidemic curve.

For these reasons, the model begins with an alert system that detects the early phases of epidemic acceleration from real-time hospitalization data. This alert system enables the integration of mechanistic SIR dynamics with historical trends, allowing peak forecasts to evolve adaptively as information accumulates during the ascending phase of the epidemic.

\paragraph{Data Preprocessing and Curve Smoothing.} Hospitalization counts are recorded daily at each pediatric center. For each day $t$, a 15-day local window (7 days before and 7 days after) is used to apply a preliminary moving-average smoothing step. This step reduces daily reporting variability and prepares the time series for subsequent smoothing. 

To obtain a differentiable representation of the incidence curve, a Savitzky–Golay filter is applied repeatedly—under slightly varied numerical configurations—to stabilize the estimation of higher-order derivatives. These realizations are then aggregated to produce a smooth and robust representation of the daily hospitalization curve, which is updated sequentially as new daily observations become available. The resulting continuous curve is denoted by $H(t)$ for $t > 0$.

\paragraph{Three-Level Alert System.} From the smoothed curve $H(t)$, we extract three epidemiologically meaningful time points:

\begin{itemize}
\item $\mathbf{t}_{0}$: Onset of growth, defined as the earliest time at which $dH>0$ and $d^{2}H>0$.
\item $\mathbf{t}_{\min}$: Acceleration alert, defined as the moment when the second derivative of $H$ reaches its highest value while the curve is still increasing ($dH>0$), corresponding to the point of maximum concavity during the ascending phase.
\item $\mathbf{t}_{1}$: Inflection alert, the classical epidemic inflection point where $d^{2}H = 0$ and the growth rate of hospitalizations reaches its maximum.
\end{itemize}

\begin{rmk}
    Here, $dH$ and $d^2 H$ denote the first and second derivatives with respect to time of the smoothed daily hospitalization curve $H(t)$, and are used to characterize changes in growth rate and curvature of the epidemic trajectory.
\end{rmk}

The triplet $(\mathbf{t}_{0}, \mathbf{t}_{\min}, \mathbf{t}_{1})$ forms the core of the alert system.
The alerts $\mathbf{t}_{\min}$ and $\mathbf{t}_{1}$ serve as the key triggers for peak-date and peak-magnitude predictions, respectively. A threshold $\mu$ is imposed for $\mathbf{t}_{0}$ to prevent activation during periods of anomalously low incidence, ensuring model stability in atypical seasons.

\paragraph{Mechanistic Component: Seasonal SIR Mode.} To inform forecasts during the pre-peak period, we employ the seasonal Susceptible–Infectious–Recovered (SIR) framework proposed by Weber \cite{weber2001}. 

The model is defined by the following system of ordinary differential equations (ODEs):
\begin{equation}\label{SIR2}
    \begin{cases}
      \cfrac{d S}{d t}(t) = - \beta(t) S(t) I(t) + \gamma R(t),\\
      \cfrac{d I}{d t}(t) = \beta(t) S(t) I(t) - \nu I(t), \\
      \cfrac{d R}{d t}(t) = \nu I(t) - \gamma R(t),
    \end{cases}\,
\end{equation}
where $S(t)$, $I(t)$, and $R(t)$ denote the proportions of the population that are susceptible, infectious, and recovered at time $t$, respectively, satisfying $S+I+R=1$.

The transmission rate is modeled as a seasonal function,
$$\beta(t) = b_0 \cdot \left(1 + b_1 \cdot \cos(2\pi t + \phi)\right) \quad t>0,$$
which captures the strong annual seasonality observed in pediatric respiratory infections.

In order to link the epidemiological dynamics to observable healthcare demand, hospitalizations are assumed to be proportional to the prevalence of infectious individuals. Accordingly, the SIR-based forecast of daily hospitalization incidence is given by
$$H_{\text{SIR}}(t) = \alpha \cdot I(t),\quad t>0,$$
where $\alpha$ is a scaling parameter that maps the proportion of infectious individuals to the expected number of daily hospital admissions.

The model parameters are
$$\theta=(b_0, b_1, \phi, \alpha, I_0, R_0),$$
where the assumed values and admissible ranges for each parameter are summarized in Table \ref{sir_constants}. %nombre de la tabla

\begin{rmk}
This seasonal SIR formulation was originally developed to describe the dynamics of respiratory syncytial virus (RSV) and is particularly well suited for pediatric respiratory infections. In this study, RSV-based dynamics are adopted due to the high prevalence of RSV among pediatric acute respiratory infections and because the available hospitalization data do not distinguish etiological agents. Consequently, the model represents the aggregate respiratory burden rather than virus-specific transmission, while retaining a biologically meaningful seasonal structure.    
\end{rmk}

\begin{table}[width=.9\linewidth,cols=4]
\caption{Parameters of the SIR model applied to RSV dynamics. The table presents each parameter, its definition, and the corresponding value or range used in this study. Constants were based on \cite{weber2001} for RSV-specific dynamics.}\label{sir_constants}
\begin{tabular*}{\tblwidth}{@{} LLLL@{} }
\toprule
Parameter& Definition & Value/Range  \\
\midrule
$S_0 =S(0)$ & Initial susceptible rate & $1-I_0-R_0$  \\
$I_0 =I(0)$ & Initial infected rate & $ [0, 0.5]$ \\
$R_0 =R(0)$ & Initial recovered rate & $ [0, 0.5]$  \\
$b_0$ & Mean transmission rate & $ (0, 3000]$  \\
$b_1$ & Seasonal amplitude & $ [0, 1]$  \\
$\phi$ & Seasonal forcing & $ [0, 2\pi]$ \\
$\alpha$ & hospitalization rate & $(0, 2000]$  \\
$\gamma$ & Loss-of-immunity rate & 1.8 \\
$\nu$ & Recovery rate & 36  \\
\bottomrule
\end{tabular*}
\end{table}

\begin{figure}
	\centering
		\includegraphics[scale=.28]{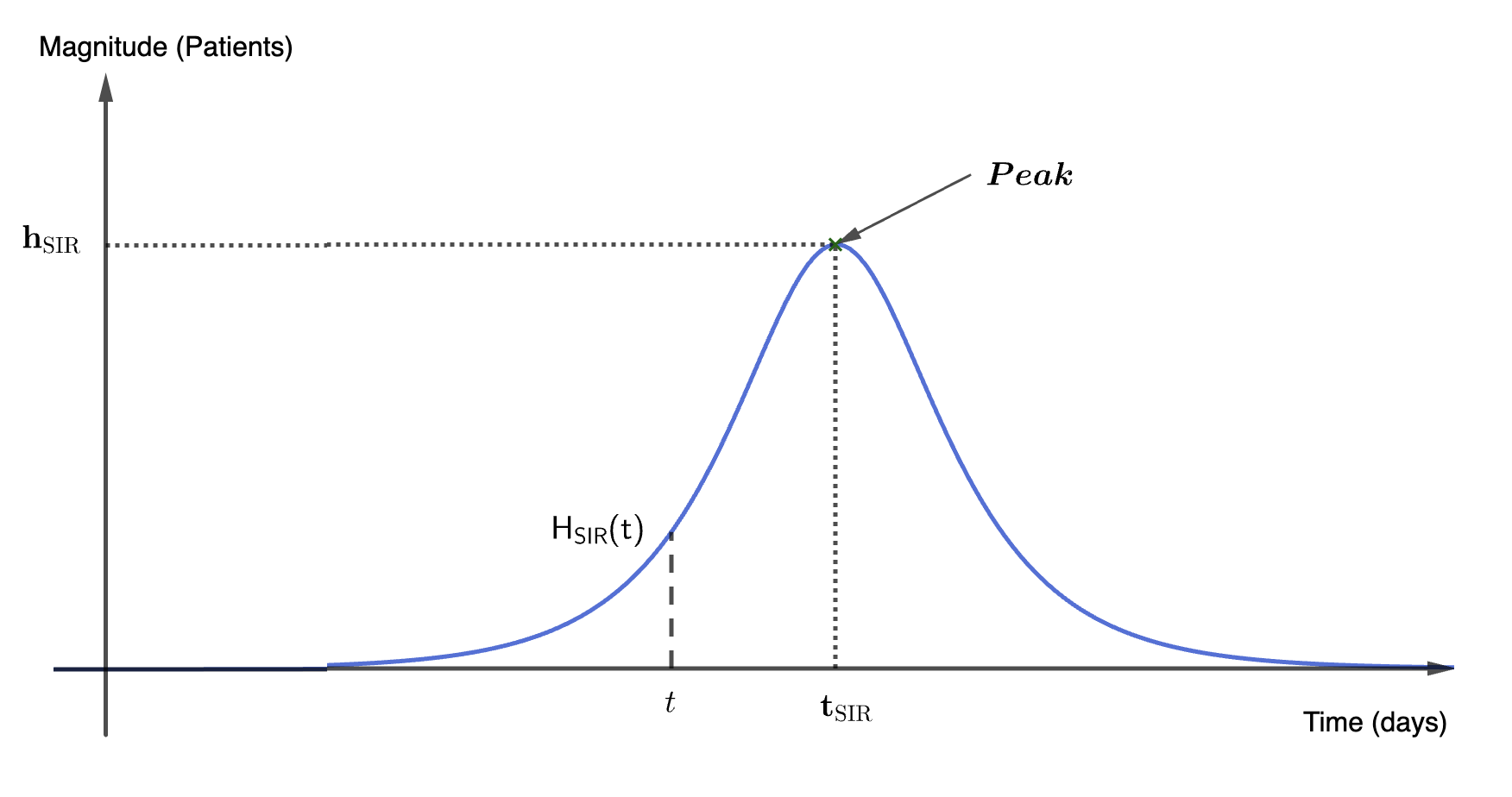}
	\caption{SIR-based hospitalization curve $H_{\text{SIR}}(t)$ showing the predicted peak magnitude $\mathbf{h}_{\text{SIR}}$ and its corresponding peak day $\mathbf{t}_{\text{SIR}}$.}
	\label{FIG:sir_curve}
\end{figure}

Given $\theta$ and data available at day $t > \mathbf{t}_{\min}$, the SIR-derived peak timing and magnitude are defined as (see Figure \ref{FIG:sir_curve}):
$$\mathbf{t}_{\text{SIR}} (t, \theta) = \text{arg} \max_{\tau} H_{\text{SIR}}(\tau, \theta),$$ $$\mathbf{h}_{\text{SIR}} (t, \theta) = \max_{\tau} H_{\text{SIR}}(\tau, \theta).$$

Near the peak, the SIR curve provides an excellent local approximation to observed dynamics, but earlier in the epidemic season—when only partial information is available—the mechanistic model alone is insufficient for accurate long-range forecasting.

\paragraph{Historical Mobile Prediction.} To enhance anticipation capability more than 30 days before the peak, we incorporate a historical mobile prediction, defined for $t>\mathbf{t}_{\min}$ as:
$$\mathbf{t}_{M} (t) = \mathbf{t}_{\min} + \cfrac{1}{|\mathcal{Y}|} \sum_{y \in \mathcal{Y}} \left( \mathbf{t}_{\text{peak}}^{y} - \mathbf{t}_{\min}^{y} \right),$$
where $\mathcal{Y}$ is the set of historical years with available data, $\mathbf{t}_{\text{peak}}^{y}$ denotes the peak day of hospitalizations in year $y$, and $\mathbf{t}_{\min}^{y}$ denotes the corresponding acceleration alert day in that same year.

This contributes long-term epidemiological information that the SIR model alone cannot infer from early-season observations.  

\paragraph{Combined Peak-Date Forecast.} The final prediction for the peak date is a weighted combination of the historical estimator and the SIR estimator:
\begin{equation}\label{pred_fecha}
\hat{\mathbf{t}}(t) = \omega (t) \cdot \mathbf{t}_{M} (t) + (1- \omega(t)) \cdot\mathbf{t}_{\text{SIR}} (t),
\end{equation}
where the weight is defined as
$$\omega(t) = \min\left(1, \frac{t - \mathbf{t}_{\text{min}}}{\mathbf{t}_{\text{max}} - \mathbf{t}_{\text{min}}} \right),$$
and $\mathbf{t}_{\max}$ denotes the historical mean peak date. 

In this way, the historical mobile prediction has a stronger influence immediately after the alert $\mathbf{t}_{\min}$, but its contribution gradually decreases over time, so that $\hat{\mathbf{t}} \rightarrow \mathbf{t}_{\text{SIR}}$ when $t \rightarrow \mathbf{t}_{\max}$, reflecting that the SIR model remains the primary driver of the forecast as the peak approaches.
Thus the model progressively transitions from historical guidance to mechanistic inference as the season approaches its peak.

\paragraph{Parameter Estimation.} The parameter vector $\theta$ is calibrated through the loss function
\begin{multline}\label{eq:loss}
\mathcal{L}(\theta) = 
\lambda \left[ \mathrm{MSE}\!\left( H , H_{\text{SIR}}( \theta) \right)
+ \rho \left| \mathbf{h}_{\text{SIR}}( \theta) - \mathbf{h}_{0} \right|^2 \right] \\
+ (1-\lambda)\left| \mathbf{t}_{\text{SIR}} ( \theta) - \mathbf{t}_{M} \right|^2
\end{multline}

where $\mathbf{h}_{0}$ is the historical mean peak magnitude, $\rho$ penalizes unrealistic heights, and $\lambda$ controls the balance between curve-fitting and peak-alignment objectives.
Details of hyperparameter selection appear in the Appendix \ref{appendix}.

\paragraph{Peak-Magnitude Forecast.} After the inflection alert $\mathbf{t}_{1}$, the SIR approximation is sufficiently stable to provide a model-based magnitude estimate:
\begin{equation}
\hat{\mathbf{h}}(t) = \mathbf{h}_{\text{SIR}} (t) \quad t>\mathbf{t}_1.
\end{equation}
Waiting until the curve has passed its inflection avoids noisy or highly variable magnitude estimates.

\paragraph{Uncertainty Quantification.} To support operational decision-making, both forecasts are reported with interpretable uncertainty intervals:
\begin{itemize}
\item The uncertainty of $\hat{\mathbf{t}}$ is based on the historical standard deviation of $\mathbf{t}_{M}$.
\item The uncertainty of $\hat{\mathbf{h}}$ is based on the mean absolute deviation between $H_{\text{SIR}}$ and $H$, and is only displayed after $t>\mathbf{t}_{1}$.
\end{itemize}

\begin{figure*}
	\centering
		\includegraphics[width=\textwidth]{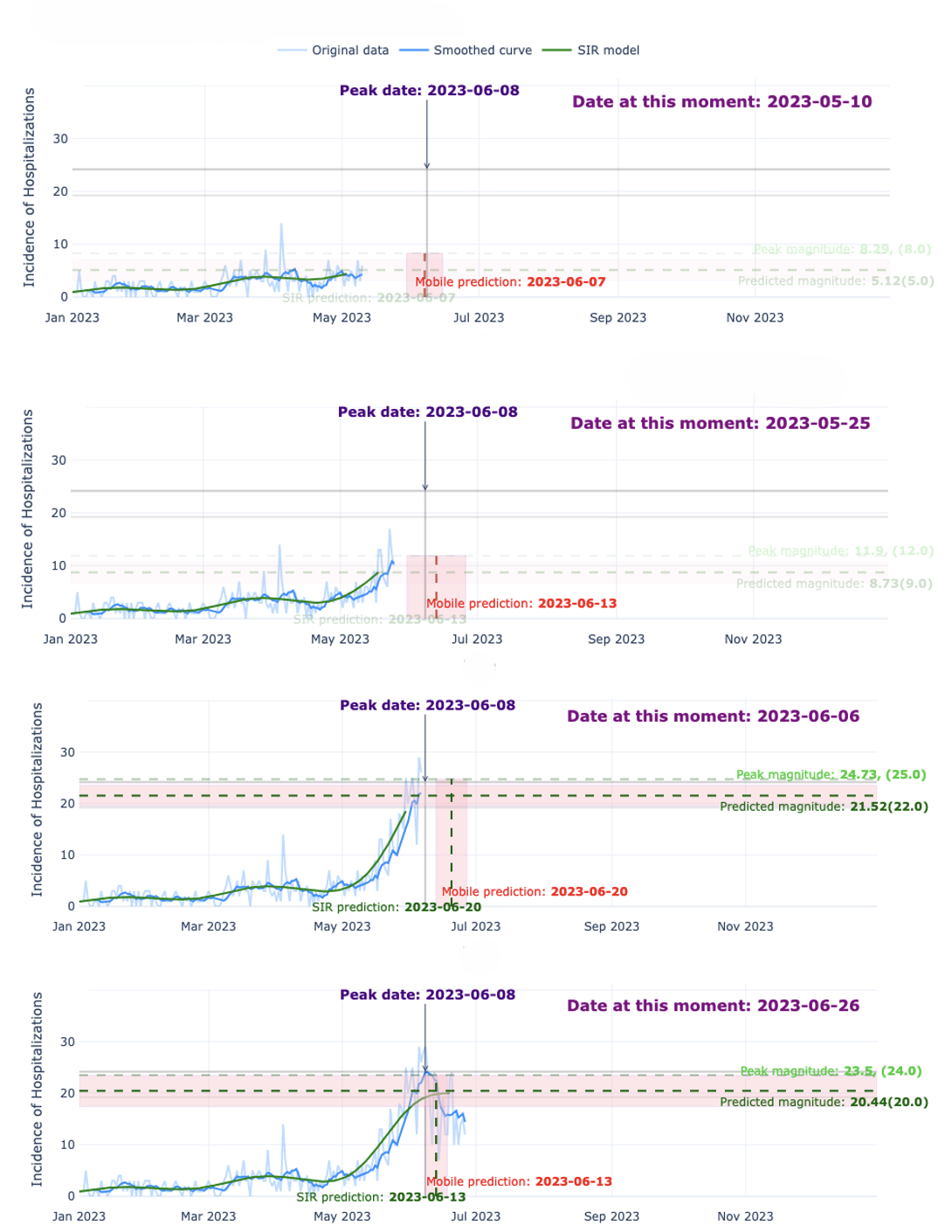}
	\caption{Evolution of the epidemiological curve during May–June 2023 at four selected dates (i: May 10, ii: May 25, iii: June 6, iv: June 26). Each panel shows the original data (light blue), the smoothed curve (blue), and the SIR model (green). Vertical markers indicate the predicted peak date with its range, while horizontal markers denote the predicted peak magnitude with its range. The x-axis represents time (date) and the y-axis the incidence magnitude.}
	\label{FIG:HCLM_2023_evol}
\end{figure*}

Figure \ref{FIG:HCLM_2023_evol} illustrates the temporal flow of the alert system within the hospital setting (for the case of HLCM in 2023), and Algorithm \ref{alg_modelo} summarizes the predictive pipeline.

\begin{algorithm}\label{alg_modelo}
\caption{Alert-Based Forecasting of Peak Date and Magnitude}
\label{alg:alert-peak}
\DontPrintSemicolon
\KwIn{Current day $t$. }
\KwOut{ $\hat{\mathbf{t}}(t)$, $\hat{\mathbf{h}}(t)$ (if available), and uncertainty ranges}

\textbf{Params:} alert threshold $\mu$; SIR parameter bounds; loss weights $\lambda$ and $\rho$.\;

\BlankLine
\textbf{Step 1. Preprocessing.}\;
Use local moving average and the Savitzky--Golay filter to obtain $H$.\;

\BlankLine
\textbf{Step 2. Alert system.}\;
Identify $\mathbf{t}_0$ as the first time such that $H'>0$, $H''>0$ and $H>\mu$.\;
\If{$t < \mathbf{t}_0$}{\Return{no alert / prediction not available.}}
Identify $\mathbf{t}_{\min}$ (acceleration alert) and $\mathbf{t}_1$ (inflection point / maximum growth rate).\;
\If{$t < \mathbf{t}_{\min}$}{\Return{alert not yet triggered / prediction not available.}}

\BlankLine
\textbf{Step 3. Mobile prediction.}\;
Compute the mobile prediction $\mathbf{t}_{M}(t)$.

\BlankLine
\textbf{Step 4. SIR calibration.}\;
Fit the seasonal SIR model with parameter vector $\theta = (b_0, b_1, \phi, \alpha, I_0, R_0)$
by minimizing the loss function $\mathcal{L}(\theta)$. \; Let $\theta^\star$ denote the minimizer.\;
From the SIR curve $H_{\text{SIR}}(t,\theta^\star)$, compute
$\mathbf{t}_{\text{SIR}}(t) = \mathbf{t}_{\text{SIR}}(t,\theta^\star)$ and  
$\mathbf{h}_{\text{SIR}}(t) = \mathbf{h}_{\text{SIR}}(t,\theta^\star)$.\;

\BlankLine
\textbf{Step 5. Combined peak-date prediction.}\;
Compute the weight $\omega(t)$. Set the combined peak-date prediction as
\[
\hat{\mathbf{t}}(t)
= \omega(t)\,\mathbf{t}_{M}(t)
+ \bigl[1-\omega(t)\bigr]\,\mathbf{t}_{\text{SIR}}(t).
\]

\BlankLine
\textbf{Step 6. Peak-magnitude prediction.}\;
\If{$t > \mathbf{t}_1$}{
    Set $\hat{\mathbf{h}}(t) = \mathbf{h}_{\text{SIR}}(t)$.\;
}
\Else{
    Set $\hat{\mathbf{h}}(t)$ as unavailable (magnitude prediction deferred until after $\mathbf{t}_1$).\;
}

\BlankLine
\textbf{Step 7. Uncertainty ranges.}\;
Construct a date interval around $\hat{\mathbf{t}}(t)$.\;
If $t > \mathbf{t}_1$, construct a magnitude interval around $\hat{\mathbf{h}}(t)$.\;

\BlankLine
\textbf{Step 8. Output.}\;
\Return $\hat{\mathbf{t}}(t)$, $\hat{\mathbf{h}}(t)$ (if available), and their associated uncertainty ranges.\;

\end{algorithm}

%%%%%%%%%%%%%%%%%%%%%%%%%%%%%%%%%%%%%%%%%%%%%%%%%%%%%%%%%%%%%%%%%%%%%%%%%%%%%%%%%%%%%%%%%%%%%%%%%%

\section{Evaluation and Results}\label{evaluation}

The evaluation of the proposed predictive framework was designed to mirror its intended operational use: generating forecasts of peak date and peak magnitude {\bf before} the true epidemic peak occurs. Although the assessment is performed retrospectively—using past seasons as test cases—the methodology strictly adheres to a real-time forecasting perspective. Predictions are produced sequentially from the moment the alert system activates, using only information available up to that day. This prevents any leakage of future data: unlike retrospective curve fitting, forecasts do not automatically converge to the true peak after its occurrence.

\paragraph{Evolution of Predictions During the Epidemic Season.} Figure~\ref{FIG:HCLM_2023_evol} illustrates the evolution of the epidemiological curve for HLCM during May–June 2023 at four selected monitoring days. Each panel shows the raw daily hospitalization incidence data (light blue), the smoothed incidence curve (blue), and the fitted SIR curve (green). Vertical markers denote the predicted peak date and its uncertainty range, while horizontal markers indicate the predicted peak magnitude and its corresponding interval.

This sequence highlights how predictions evolve as new data become available: early in the season, forecasts are guided primarily by historical trends and the initial alert signals; later, once the inflection point is reached, the SIR component dominates and predictions stabilize.

A key feature visible in Figure~\ref{FIG:HCLM_2023_evol} is the method’s {\bf robustness to short-term fluctuations}. Pediatric hospitalization data often exhibit substantial noise, and the maximum of the smoothed curve may not coincide with isolated spikes in the raw series. The model is specifically designed to prioritize the operational peak—the sustained period of maximum healthcare pressure—rather than transient anomalies caused by reporting inconsistencies or atypical events. This emphasis on persistence rather than isolated spikes is essential for hospital planning and aligns with real-world definitions of system saturation.

\begin{figure*}
	\centering
		\includegraphics[width=\textwidth]{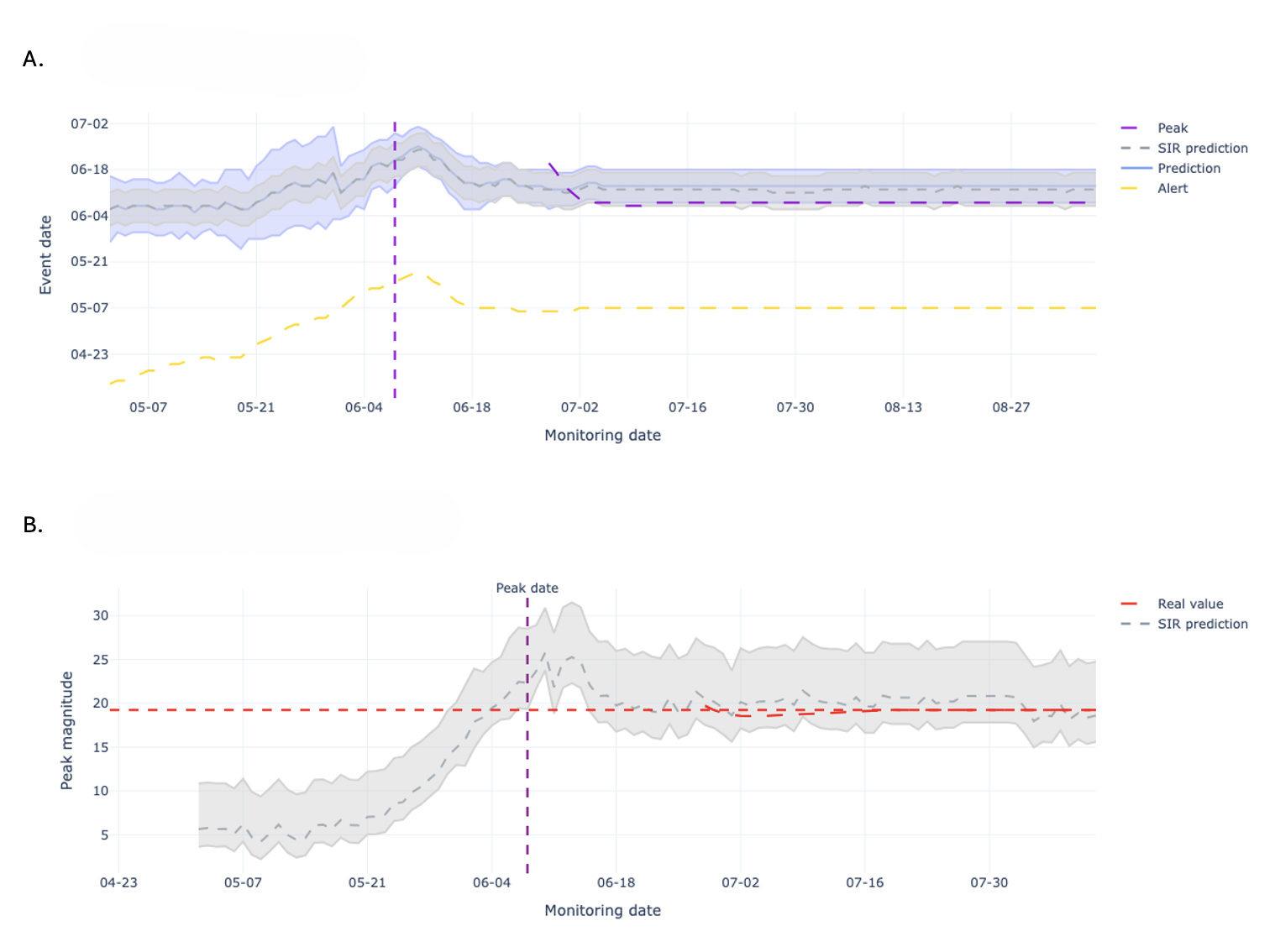}
	\caption{Predictive performance of the ensemble model for HLCM during the 2023 season. (A) Evolution of peak date forecasts over time. The x-axis shows the monitoring date, while the y-axis indicates the predicted peak date. Gray denotes the SIR-based prediction, blue the mobile prediction, yellow the alert signal, and purple the observed peak. The vertical and horizontal markers illustrate the stabilization of the alert system. (B) Evolution of peak magnitude forecasts. The x-axis represents the monitoring date, and the y-axis the predicted peak magnitude. The ensemble prediction is shown in gray, and the observed peak magnitude in red}
	\label{FIG:hlcm2023}
\end{figure*}

\begin{figure*}
	\centering
		\includegraphics[width=\textwidth]{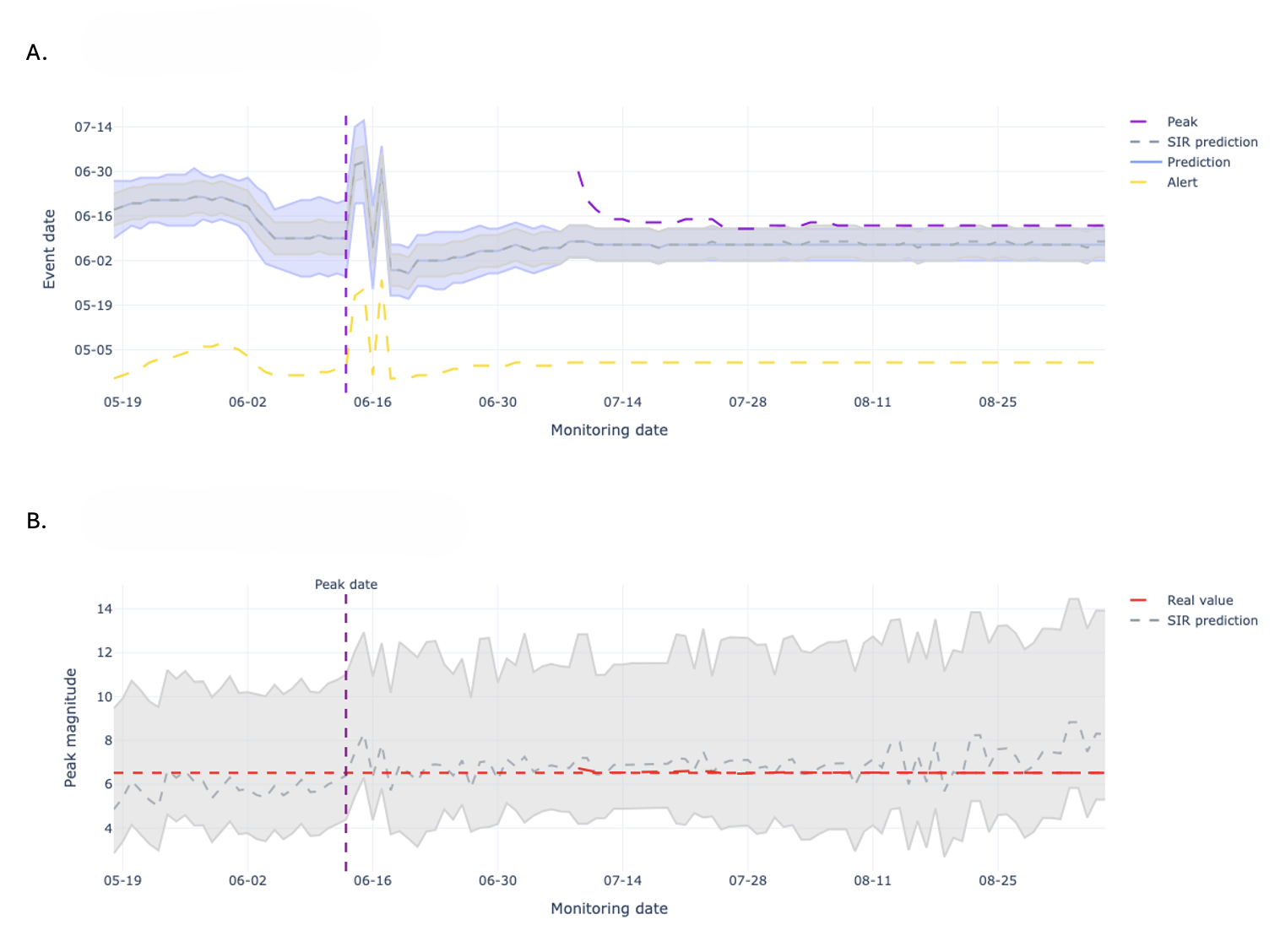}
	\caption{Predictive performance of the ensemble model for HLCM during the 2024 season. (A) Evolution of peak date forecasts over time. The x-axis shows the monitoring date, while the y-axis indicates the predicted peak date. Gray denotes the SIR-based prediction, blue the mobile prediction, yellow the alert signal, and purple the observed peak. The vertical and horizontal markers illustrate the stabilization of the alert system. (B) Evolution of peak magnitude forecasts. The x-axis represents the monitoring date, and the y-axis the predicted peak magnitude. The ensemble prediction is shown in gray, and the observed peak magnitude in red}
	\label{FIG:hlcm2024}
\end{figure*}

\paragraph{Daily Evolution of Peak Predictions.} Figures~\ref{FIG:hlcm2023} and~\ref{FIG:hlcm2024} present the day-by-day evolution of peak date and magnitude forecasts for HLCM during the 2023 and 2024 seasons, respectively.

\begin{itemize}
    \item Panel A in each figure displays the evolution of the peak date prediction, where the x-axis corresponds to the monitoring day and the y-axis to the predicted peak date.
    
    The SIR-based estimate is shown in gray, the mobile historical prediction in blue, the alert signals in yellow, and the observed peak in purple. Vertical and horizontal markers illustrate the stabilization of the alert system and the convergence of the ensemble forecast.
    
    \item Panel B shows the evolution of the peak magnitude prediction, where the y-axis indicates the predicted peak magnitude. The ensemble prediction appears in gray, and the observed magnitude is marked in red.
\end{itemize}

Across both years, predictions become progressively precise as the season unfolds. Early forecasts provide a wide but meaningful planning window, while later predictions exhibit narrower uncertainty intervals. This progressive narrowing is a cornerstone of the framework: peak date predictions offer the longest anticipation window, whereas magnitude predictions become reliable only after the epidemic’s acceleration and inflection alerts have been triggered.

\paragraph{Anticipation of Epidemic Peaks.} Table~\ref{results_anticip} summarizes the anticipation performance of the model across all four hospitals during the 2023–2024 seasons. Predictions are evaluated at three operationally relevant anticipation windows: 1 month prior to the observed peak, 2 weeks prior, and 1 week prior.

For each hospital and season, the table reports the correctness of both the peak date and magnitude predictions, indicated with a color scale: green for accurate forecasts, yellow for near misses, and red for unsuccessful predictions. This categorization offers an intuitive summary of operational usefulness, capturing the ability of the model to generate timely and actionable alerts.

The results highlight that:

\begin{itemize}
    \item All hospitals achieved meaningful anticipation of peak timing at least two weeks before the observed peaks.
    \item The acceleration alert $\mathbf{t}_{\min}$ consistently activated early, enabling forecasts with substantial lead time.
    \item Peak magnitude predictions were consistently reliable once $t > \mathbf{t}_{1}$, reflecting the stabilizing influence of the SIR component
\end{itemize}

\paragraph{Quantitative Accuracy Across Years and Hospitals.} To complement the anticipation analysis, Table~\ref{tab:metrics_summary} provides quantitative metrics for all available seasons from 2017 to 2024 (excluding 2020–2021). For each hospital, the table reports:

\begin{itemize}
    \item Anticipation in days (difference between forecast and true peak date at the time predictions stabilized),
    \item Peak date error (predicted minus observed, in days),
    \item Peak magnitude error (difference in the number of cases of the smoothed curve),
    \item Mean and standard deviation across years.
\end{itemize}

\begin{rmk}
All comparisons in Table~\ref{tab:metrics_summary} are performed with respect to the peak of the \emph{smoothed} hospitalization curve. The true peak date and magnitude are therefore defined as the maximum of the stabilized incidence curve, rather than the raw daily series, which may contain isolated fluctuations not representative of sustained hospital system pressure.
\end{rmk}

These results show moderate year-to-year variability —reflecting inherent epidemiological fluctuations— but demonstrate stable performance across hospitals and seasons. Importantly, the peak date error remains within an operationally acceptable range, particularly considering that magnitude predictions only stabilize after the inflection alert.

\paragraph{Summary of Predictive Performance.} Taken together, Figures~\ref{FIG:HCLM_2023_evol},~\ref{FIG:hlcm2023},~\ref{FIG:hlcm2024}, and Tables~\ref{results_anticip}–\ref{tab:metrics_summary} provide a comprehensive evaluation of the model’s forecasting capacity. The results demonstrate that:

\begin{itemize}
    \item The alert system reliably identifies the onset and acceleration of epidemic growth.
    \item The dual-component model (historical + SIR) offers robust early predictions with increasing precision closer to the peak.
    \item The framework provides operationally meaningful anticipation, both for peak date (long horizon) and peak magnitude (shorter horizon).
    \item The model performs consistently across multiple hospitals and epidemic seasons, supporting its practical utility for healthcare resource planning in Chile.
\end{itemize}

\begin{table*}%[htbp]
\centering
\scriptsize
\setlength{\tabcolsep}{4pt}
\renewcommand{\arraystretch}{1.2}
\caption{Anticipation of epidemic peaks by hospital during 2023–2024. The table reports relative peak values for date and magnitude across HLCM, HEGC, HFB, and HDRD. Predictions are evaluated retrospectively at three anticipation windows—1 month, 2 weeks, and 1 week prior to the observed peak—showing both alert generation and peak forecasts. Correct predictions are highlighted in green, near misses in yellow, and failed predictions in red.}
\label{results_anticip}
\begin{tabularx}{\textwidth}{l
    >{\centering\arraybackslash}X
    >{\centering\arraybackslash}X
    >{\centering\arraybackslash}X
    >{\centering\arraybackslash}X
    >{\centering\arraybackslash}X
    >{\centering\arraybackslash}X
    >{\centering\arraybackslash}X
    >{\centering\arraybackslash}X}
\toprule
\multirow{2}{*}{\textbf{Hospital}} &
\multicolumn{2}{c}{\textbf{Real}} &
\multirow{2}{*}{\textbf{Alert}} &
\multirow{2}{*}{\makecell{\textbf{1 month before}\\ \textbf{the peak}}} &
\multicolumn{2}{c}{\textbf{2 weeks before the peak}} &
\multicolumn{2}{c}{\textbf{1 week before the peak}} \\
\cmidrule(lr){2-3} \cmidrule(lr){6-7} \cmidrule(lr){8-9}
& \textbf{Peak date} & \textbf{Peak magnitude} & &
\textbf{Predicted date} &
\makecell{\textbf{Predicted}\\\textbf{date}} &
\makecell{\textbf{Predicted}\\\textbf{magnitude}} &
\makecell{\textbf{Predicted}\\\textbf{date}} &
\makecell{\textbf{Predicted}\\\textbf{magnitude}} \\
\midrule
\multicolumn{9}{c}{\textbf{ Year: 2023}} \\
\midrule
HLCM & June 8th  & 20 & May 7th  &
\cellcolor{green!35}{June 6th $\pm$ 8 days} &
\cellcolor{green!35}{June 13th $\pm$ 13 days} &
\cellcolor{red!10}{10 $\pm$ 2 people} &
\cellcolor{green!35}{June 11th $\pm$ 8 days} &
\cellcolor{yellow!5}{16 $\pm$ 2 people} \\
HEGC & June 10th & 20 & May 8th  &
\cellcolor{green!35}{June 9th $\pm$ 10 days} &
\cellcolor{green!35}{June 5th $\pm$ 10 days} &
\cellcolor{red!10}{9 $\pm$ 3 people} &
\cellcolor{yellow!5}{June 20th $\pm$ 6 days} &
\cellcolor{yellow!5}{17 $\pm$ 2 people} \\
HFB  & June 15th & 13 & May 15th &
-- &
\cellcolor{green!35}{June 24th $\pm$ 9 days} &
\cellcolor{red!10}{4 $\pm$ 2 people} &
\cellcolor{yellow!5}{June 29th $\pm$ 12 days} &
\cellcolor{green!35}{13 $\pm$ 2 people} \\
HRDR & June 21st & 11 & April 12th &
\cellcolor{yellow!5}{June 2nd $\pm$ 15 days} &
\cellcolor{red!10}{May 25th $\pm$ 15 days} &
\cellcolor{yellow!5}{8 $\pm$ 2 people} &
\cellcolor{red!10}{May 28th $\pm$ 12 days} &
\cellcolor{green!35}{9 $\pm$ 3 people} \\
\midrule
\multicolumn{9}{c}{\textbf{Year: 2024}} \\
\midrule
HLCM & June 14th & 7  & May 6th  &
-- &
\cellcolor{green!35}{June 21st $\pm$ 7 days} &
-- &
\cellcolor{green!35}{June 9th $\pm$ 11 days} &
\cellcolor{green!35}{6 $\pm$ 2 people} \\
HEGC & June 26th & 10 & May 22nd &
-- &
\cellcolor{red!10}{July 17th $\pm$ 9 days} &
-- &
\cellcolor{red!10}{July 17th $\pm$ 9 days} &
-- \\
HFB  & --        & -- & --       &
-- & -- & -- & -- & -- \\
HRDR & June 25th & 10 & May 21st &
-- &
\cellcolor{red!10}{July 18th $\pm$ 12 days} &
\cellcolor{green!35}{10 $\pm$ 4 people} &
\cellcolor{red!10}{July 22nd $\pm$ 15 days} &
\cellcolor{yellow!5}{6 $\pm$ 3 people} \\
\bottomrule
\end{tabularx}
\end{table*}

\begin{table*}%[htbp]
\centering
\small
\setlength{\tabcolsep}{3pt}
\renewcommand{\arraystretch}{1.2}
\caption{Anticipation, peak date error, and peak magnitude error by year and hospital (2017–2024). The table summarizes performance metrics for each hospital (HLCM, HEGC, HFB, HDRD), excluding 2020–2021. For each year, the anticipation in days, the error in peak date (days), and the error in peak magnitude (number of cases of the smoothed curve) are reported. Mean and standard deviation values are also provided, allowing comparison of variability across hospitals and years.}
\label{tab:metrics_summary}
\begin{tabularx}{\linewidth}{
    >{\centering\arraybackslash}p{0.08\linewidth} % Year
    >{\raggedright\arraybackslash}p{0.30\linewidth} % Metric (wrap)
    >{\centering\arraybackslash}p{0.12\linewidth}  % HLCM
    >{\centering\arraybackslash}p{0.12\linewidth}  % HEGC
    >{\centering\arraybackslash}p{0.12\linewidth}  % HFB
    >{\centering\arraybackslash}p{0.12\linewidth}  % HRDR
}
\toprule
 & \textbf{Feature} & \textbf{HLCM} & \textbf{HEGC} & \textbf{HFB} & \textbf{HRDR} \\
\midrule
\multirow{3}{*}{2017}
& Anticipation (days)         & 39  & 36  & 45  & 66 \\
& peak date error (days)      & 7   & 0   & 12  & 7  \\
& peak magnitude error (people) & 7.53 & 12.13 & 7.76 & 12.95 \\
\midrule
\multirow{3}{*}{2018}
& Anticipation (days)         & 31  & 42  & 46  & 32 \\
& peak date error (days)      & 13  & 0   & 7   & 26 \\
& peak magnitude error (people) & 8.87 & 7.30 & 5.75 & 10.36 \\
\midrule
\multirow{3}{*}{2019}
& Anticipation (days)         & 47  & 59  & 42  & 35 \\
& peak date error (days)      & 30  & 20  & 5   & 17 \\
& peak magnitude error (people) & 11.10 & 11.33 & 12.93 & 10.33 \\
\midrule
\multirow{3}{*}{2022}
& Anticipation (days)         & 26  & 43  & 20  & 48 \\
& peak date error (days)      & 6   & 6   & 11  & 4  \\
& peak magnitude error (people) & 2.72 & 7.08 & 0.70 & 10.45 \\
\midrule
\multirow{3}{*}{2023}
& Anticipation (days)         & 37  & 40  & 27  & 55 \\
& peak date error (days)      & 6   & 4   & 10  & 5  \\
& peak magnitude error (people) & 14.46 & 13.55 & 6.20 & 3.55 \\
\midrule
\multirow{3}{*}{2024}
& Anticipation (days)         & 26  & 19  & --  & 22 \\
& peak date error (days)      & 2   & 21  & --  & 31 \\
& peak magnitude error (people) & 1.13 & 5.15 & --  & 0.85 \\
\midrule
\multirow{3}{*}{Mean}
& Anticipation (days)         & 34.3333 & 39.8333 & 36.0000 & 43.0000 \\
& peak date error (days)      & 10.6667 & 8.5000  & 9.0000  & 15.0000 \\
& peak magnitude error (people) & 7.6350  & 9.4233  & 6.6680  & 8.0817 \\
\midrule
\multirow{3}{*}{Std}
& Anticipation (days)         & 8.2381  & 12.8906 & 11.7686 & 16.2727 \\
& peak date error (days)      & 10.1127 & 9.5864  & 2.9155  & 11.5412 \\
& peak magnitude error (people) & 5.0318  & 3.3542  & 4.3899  & 4.7411 \\
\bottomrule
\end{tabularx}
\end{table*}

%------------------------
\section{Discussion}\label{discusion}

The results presented in the previous section demonstrate both the strengths and limitations of the proposed alert-based forecasting framework for anticipating pediatric ARI hospitalization peaks. The model’s behavior, illustrated in Figure~\ref{FIG:HCLM_2023_evol}, highlights its adaptive nature: although the seasonal SIR component assumes fixed parameters, these parameters are recalibrated as new data accumulate. Their fixed interpretation only becomes meaningful once the epidemic wave has fully developed, which explains why forecasts evolve throughout the season. For RSV, the infection-specific parameters reported in \cite{weber2001} were retained, while population-dependent parameters—affected by factors such as hospital type, regional characteristics, and age distribution—were found to vary substantially among years. This variability motivated the model’s adaptive calibration strategy.

A central finding of this study concerns the evolution of prediction intervals for peak date and magnitude. As shown in Table~\ref{results_anticip}, prediction intervals become operationally informative only in the days leading up to the peak. While the model reliably identifies the timing of the peak well in advance, the magnitude of the peak tends to stabilize only within a narrower anticipation window—typically around seven to ten days before the observed maximum. This observation directly influenced how predictions were communicated to hospital personnel: peak date forecasts were provided as soon as the acceleration alert $\mathbf{t}_{\min}$ was triggered, often more than one month before the peak, whereas magnitude predictions were shared only once they achieved operational stability, generally one week before the peak. Clinicians involved in the model’s co-design expressed a preference for fewer but more reliable magnitude updates rather than noisy daily fluctuations, reinforcing the practical importance of this decision.

The nature of the data itself also imposes constraints. Daily pediatric hospitalization records are inherently noisy, with abrupt directional changes even after smoothing. This volatility required the introduction of a penalty term in the optimization procedure to prevent the SIR fitting process from overreacting to anomalous increases in magnitude. The penalty’s purpose is consistent with the model’s operational intent: to characterize sustained periods of system saturation rather than transient anomalies. From a planning perspective, isolated spikes that exceed the predicted peak are less relevant if they do not correspond to prolonged pressure on hospital capacity. The model’s emphasis on persistence rather than instantaneous maxima aligns with how health systems experience and manage epidemic stress.

The monitoring results for HLCM illustrate these points clearly. In 2023 (Figure~\ref{FIG:hlcm2023}), peak date forecasts were consistently accurate, with the predicted date remaining within a stable range once early-season alerts were activated. Magnitude predictions, while converging later, eventually provided reliable estimates approximately one week before the peak. In 2024 (Figure~\ref{FIG:hlcm2024}), the model faced a different epidemiological profile: lower overall case counts, less pronounced seasonality, and increased noise after the peak. Despite this, date predictions remained operationally useful up to a week before the peak, and magnitude predictions were relatively stable due to the low incidence levels. Importantly, the post-peak fluctuations in 2024, while visually noticeable, did not undermine the model's anticipatory value.

The aggregated forecasts in Table~\ref{results_anticip} further contextualize these findings. One month before the peak, the model rarely generated complete predictions, reflecting the conservative design of the alert-based activation mechanism. Two weeks before the peak, date predictions were generally accurate across hospitals, but magnitude predictions remained unreliable—supporting the decision to limit magnitude dissemination to the final week. Across hospitals, The strongest results in 2023 were obtained for HLCM and HEGC, while HFB exhibited greater uncertainty and HRDR produced less precise predictions yet still maintained operationally useful early warnings. It is important to note that, in the Chilean public health context, a slightly premature but stable prediction is often more valuable for planning than a delayed but precise one, as the implementation of the contingency measures requires operational lead time.

The 2024 season presented additional challenges. The epidemiological curves at several hospitals showed atypical patterns—very low magnitudes at HFB, multiple local peaks at HLCM, and unstable growth phases at HEGC and HRDR. These irregularities may be partially explained by external interventions such as the Nirsevimab immunization campaign \cite{torres2025}. As a result, predictions were less consistent across hospitals, highlighting the sensitivity of the model to changes in disease dynamics and the need for ongoing adaptation in years with non-standard epidemic profiles.

Table~\ref{tab:metrics_summary} provides a broader, multi-season view of forecasting accuracy. While mean errors were similar across hospitals, the variability differed notably. HLCM exhibited low variability in peak date predictions but higher variability in peak magnitude, consistent with its complex seasonal patterns. Other hospitals showed wider fluctuations in anticipation windows but relatively stable magnitude errors. These differences likely reflect structural heterogeneity in catchment populations, referral flows, and hospital demand.

Overall, the results underscore the model’s primary strength: anticipating the timing of the seasonal peak with robust and clinically meaningful lead time, often more than one month before the peak and with excellent precision two weeks before the peak. By contrast, peak magnitude predictions are reliable only at shorter horizons—typically ten days before the peak, with optimal stability at one week—yet these values still offer significant operational value when combined with appropriate uncertainty ranges.

Therefore, the alert-based design, the integration of mechanistic and historical components, and the use of smoothing, penalties, and empirically chosen thresholds collectively produce forecasts that are resilient to data noise and adaptable to dynamic epidemic conditions. While magnitude predictions remain more sensitive to interannual variability, the accuracy and anticipation of peak-date forecasts make the framework particularly well suited for supporting hospital planning and resource allocation during winter ARI seasons.
%------------------------
\section{Conclusion}\label{conclusion}

This study introduced an alert-based ensemble forecasting model that combines seasonal SIR dynamics with historical hospitalization trends to predict the timing and magnitude of pediatric acute respiratory infection peaks in Santiago, Chile. Unlike conventional retrospective modeling approaches, the framework was designed from the outset as a practical operational tool, with outputs structured around predictive intervals, stability criteria, and an alert system that enables early and progressive forecasting throughout the epidemic season.

The results demonstrate the model’s strong ability to anticipate peak timing, offering reliable and actionable forecasts well in advance of the observed peak. In practice, the model provides an initial estimate of the peak date approximately one month before the event and achieves high accuracy two weeks before the peak, even under substantial noise in the hospitalization data. This long anticipation window is especially valuable for hospital resource planning, where decisions related to staffing, bed allocation, and surge preparedness require sufficient lead time.

Peak magnitude forecasting showed a different temporal behavior: predictions became operationally meaningful roughly ten days before the peak, with highly stable and accurate estimates during the final week prior to the observed maximum. This shorter anticipation window is consistent with the inherent variability of epidemic growth and the sensitivity of magnitude estimates to abrupt changes in hospitalization patterns. Nonetheless, these magnitude forecasts remain crucial for understanding the scale of expected hospital pressure, allowing clinical teams to adjust internal logistics and anticipate saturation thresholds.

The methodology also revealed important limitations. Forecast accuracy is sensitive to factors that disrupt typical epidemic trajectories, such as immunization campaigns, abrupt public health interventions, or rare environmental and social events. These exogenous shocks fall outside the model’s historical calibration and can generate atypical patterns, as observed in 2024. Such deviations underscore the importance of complementing mechanistic and historical components with additional sources of information and adaptive mechanisms for years marked by nonstandard dynamics.

Overall, the proposed model demonstrates that an alert-driven, hybrid forecasting strategy can provide a flexible and operationally useful approach for anticipating pediatric hospitalization surges. Although the present study is focused on Santiago, Chile, the methodology is not inherently location-specific and may be transferable to other regions with comparable healthcare structures, seasonal respiratory dynamics, and data availability. The proposed algorithmic framework can be adapted to forecast pediatric hospitalizations in different settings by recalibrating hospital-specific parameters and historical baselines, thereby extending its potential applicability beyond the local context. Future work should focus on incorporating exogenous covariates—such as meteorological indicators, viral co-circulation metrics, immunization coverage, and public health policy changes—to enhance robustness and support deployment in increasingly diverse epidemiological environments.

%----------------------------------------------------
\section*{Acknowledgments}

%This research was conducted as part of the project FONDEF ID23I10423, funded by the National Fund for Scientific and Technological Development (FONDEF), Chile. The authors gratefully acknowledge the financial support provided by FONDEF, as well as the collaboration of the Ministry of Health of Chile, whose contributions were essential for access to hospitalization data and the contextual understanding of public health needs. We also extend our special thanks to the staff of Hospital Luis Calvo Mackenna (HLCM), whose active participation and insights were invaluable for the development and implementation of this work.

%This work was supported by FONDEF project ID23I10423 and Centro de Modelamiento Matemático BASAL fund FB210005 for center of excellence, both from ANID-Chile. We gratefully acknowledge the collaboration of the Hospital Dr. Luis Calvo Mackenna  and the Chilean Ministry of Health, whose contributions, data access, and operational guidance were essential to the development of this research.

This work was supported by the Agencia Nacional de Investigación y Desarrollo (ANID) through ANID FONDEF ID23I10423, and by ANID BASAL FB210005, which supports the Centro de Modelamiento Matemático (CMM) as a Center of Excellence. We gratefully acknowledge the collaboration of the Hospital Dr. Luis Calvo Mackenna and the Chilean Ministry of Health (MINSAL), whose contributions, data access, and operational guidance were essential to the development of this research.

\paragraph{\bf Declaration of Generative AI and AI-Assisted Technologies in the Writing Process.}
During the preparation of this work, the authors used ChatGPT (OpenAI) to assist with improving grammar, clarity, and overall readability of the manuscript. After using this tool, the authors carefully reviewed, edited, and verified the content to ensure accuracy and appropriateness. The authors take full responsibility for the content of the published article.

%----------------------------------------------------
\bibliographystyle{cas-model2-names}

\bibliography{refs}

%\newpage
\appendix
\section{Appendix: Optimizer and loss function}\label{appendix}

During the implementation of the SIR model, various optimization algorithms were evaluated to fit the model parameters to the observed daily hospitalization data. The comparison included both local and global approaches, assessing their stability, computational efficiency, predictive capacity, and compatibility with an objective function characterized by multiple local minima \cite{difevol}.

In the reviewed literature, the calibration of SIR models and their variants is a decisive step for prediction quality. Studies such as \cite{piazzola2021, reis2016, reis2019, weber2001} emphasize the importance of accurately estimating parameters such as the transmission rate and infectious period to reproduce observed epidemic dynamics. Different numerical optimization and Bayesian approaches have been proposed, although in many cases the exact optimization algorithm is not specified, leaving open the question of the most suitable method for robust, stable results. An inadequate optimizer can lead to non-unique estimates, unstable solutions, or poor predictive performance; conversely, an effective optimizer can better capture epidemic dynamics, even under high variability.

In our case, the Differential Evolution algorithm was selected as the primary optimizer. Its global search capability avoids local optima and improves convergence in the presence of a non-convex objective function, yielding a more reliable model for early warning generation and hospital resource planning. Differential Evolution provided the best balance between robustness, predictive capacity, and control over peak behavior among the methods tested, making it the benchmark optimizer for this work. Table~\ref{tab:mixed_sir_models} summarizes the variants and possible combinations evaluated.

%%% aca va la tabla de optimizadores

\begin{table*}[htbp]
\centering
\footnotesize
\setlength{\tabcolsep}{6pt}
\renewcommand{\arraystretch}{1.2}
\caption{Exploration of alternative optimizers and loss functions for SIR model calibration. 
The loss functions tested include the \emph{peak date error} ($|\text{mobile prediction} - \text{peak SIR}|^2$), designed to penalize deviations in peak timing, and the \emph{Weighted MSE}, which applies exponential decay to emphasize recent observations. 
For each optimizer–loss combination, the table summarizes performance characteristics and the rationale for not adopting the method in the final implementation.}
\label{tab:mixed_sir_models}
\begin{tabularx}{\linewidth}{
  >{\raggedright\arraybackslash}p{0.18\linewidth} % Optimizer
  >{\raggedright\arraybackslash}p{0.22\linewidth} % Loss function
  >{\raggedright\arraybackslash}X                 % Remark (ajusta automático)
}
\toprule
\textbf{Optimizer} & \textbf{Loss function} & \textbf{Remark} \\
\midrule
\multirow{2}{*}{L-BFGS-B1} 
& MSE + peak date error & Produced magnitude stabilization but highly variable peak date estimates; not retained due to inconsistent temporal predictions. \\
& Weighted MSE + peak date error & Improved responsiveness to changes, but generated unstable peak estimates in partial series and inconsistent magnitudes in full series. \\
\midrule
Dual Annealing 
& MSE + peak date error & Achieved balanced predictions in $\sim$5s with rate $\approx$0.9, but some forecasts were not epidemiologically plausible. \\
\midrule
\multirow{2}{*}{Differential Evolution} 
& MSE + peak date error & Yielded realistic curves for partial series but produced implausible forecasts at evaluation onset. A mixed SIR approach with the full series (rate 0.95--0.99) provided the best fit, closely tracking rolling predictions, though sensitive to noise. \\
& Weighted MSE + peak date error & Generated unstable fits with substantial deviations from observed data, particularly early in the season; not retained. \\
\midrule
SHGO 
& MSE + peak date error & Produced solutions inconsistent with epidemiological dynamics; not retained. \\
\bottomrule
\end{tabularx}
\end{table*}

The optimizer was configured to minimize a loss function composed of the mean squared error (MSE) between the observed and simulated series, combined with a penalty for discrepancies in peak timing and magnitude (see Eq. \ref{eq:loss}).

\begin{table}%[width=.9\linewidth,cols=4]
\caption{Loss function parameters: selection of $\lambda$ and $\rho$. Summary of the implemented values of the penalization weights $\lambda$ (peak timing penalty) and $\rho$ (magnitude penalty) chosen for each hospital. These values correspond to the configuration adopted in the final model to balance the trade-off between accuracy in peak date and peak magnitude prediction.}\label{loss_func_par}
\begin{tabular*}{\tblwidth}{@{} LLLL@{} }
\toprule
Hospital & $\lambda$ & $\rho$ \\
\midrule
HLCM & 0.9981 & 0.97   \\
HEGC & 0.998 & 0.93   \\
HFB & 0.9991 & 0.8    \\
HRDR & 0.9991 & 0.2   \\
\bottomrule
\end{tabular*}
\end{table}

\begin{figure*}%[h!] % [h!] is a float option for "here" if possible
    \centering % Centers the image
    \includegraphics[width=\textwidth]{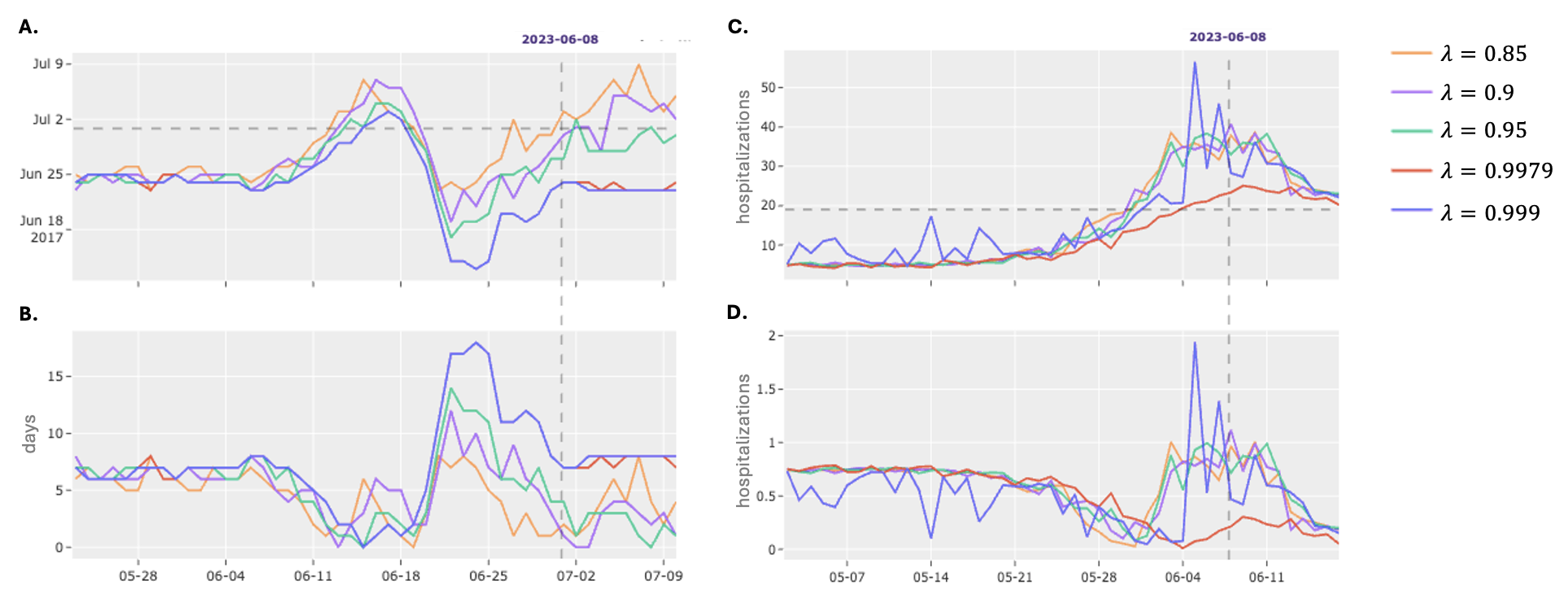} % Adjust width as needed
    \caption{Impact of varying $\lambda$ in the loss function for the HLCM case (2023). The effect of different penalization weights $\lambda$ is shown across four panels: (A) predicted peak date over monitoring days; (B) prediction error in days relative to the observed peak; (C) predicted peak magnitude (hospitalizations); and (D) magnitude error relative to observed values. The selected $\lambda$ is highlighted in red, while alternative candidate values are displayed for comparison..}
    \label{fig:lambda} % Optional: for referencing the figure
\end{figure*}

Once the loss function structure was defined, a grid search was conducted to explore different values of $\lambda$, which regulates the trade-off between peak date accuracy and overall curve fit. Figure~\ref{fig:lambda} shows the impact of varying $\lambda$ on the prediction for HLCM in 2023. Values below 0.8 degraded the quality of the date prediction, while values near 0.9 provided the best compromise between temporal accuracy and stability. Higher values $(\lambda > 0.99)$ improved magnitude accuracy but occasionally produced day-to-day outliers in predicted peak height.

To prevent unrealistic fluctuations in peak magnitude while retaining the benefits of high $\lambda$ values, two additional strategies were tested:
(a) adding a penalty term for deviations from the historical maximum peak magnitude, and
(b) applying a time-weighting scheme to the observed data.
The first approach effectively suppressed outliers without compromising fit quality; the second was discarded because it distorted the representation of the original series.

The hyperparameters $\lambda$ and $\rho$ were fine-tuned to satisfy three operational criteria:
\begin{itemize}
    \item Continuity with the observed series.
    \item Alignment of predicted and observed peaks in both magnitude and date.
    \item Stability without outlier behavior.
\end{itemize}

This calibration was computationally demanding but ensured a robust and operationally reliable configuration. Table~\ref{loss_func_par} presents the selected values for each hospital, tested across the years 2015–2023 (excluding 2020 and 2021, due to COVID-19).

\end{document}